\documentclass[essd, manuscript]{copernicus}

\nolinenumbers

\begin{document}

\title{WeatherBench: A benchmark dataset for data-driven weather forecasting}


\Author[1]{Stephan}{Rasp}
\Author[2]{Peter D.}{Dueben}
\Author[3]{Sebastian}{Scher}
\Author[4]{Jonathan A.}{Weyn}
\Author[5]{Soukayna}{Mouatadid}
\Author[1]{Nils}{Thuerey}

\affil[1]{Technical University of Munich, Germany}
\affil[2]{European Centre for Medium-range Weather Forecasts, Reading, UK}
\affil[3]{Department of Meteorology and Bolin Centre for Climate Research, Stockholm University, Sweden}
\affil[4]{Department of Atmospheric Sciences, University of Washington, Seattle, USA}
\affil[5]{Department of Computer Science, University of Toronto, Canada}



\correspondence{Stephan Rasp (stephan.rasp@tum.de)}

\runningtitle{WeatherBench}

\runningauthor{Rasp et al.}

\received{}
\pubdiscuss{} 
\revised{}
\accepted{}
\published{}


\firstpage{1}

\maketitle

\begin{abstract}
Data-driven approaches, most prominently deep learning, have become powerful tools for prediction in many domains. A natural question to ask is whether data-driven methods could also be used to predict global weather patterns days in advance. First studies show promise but the lack of a common dataset and evaluation metrics make inter-comparison between studies difficult. Here we present a benchmark dataset for data-driven medium-range weather forecasting, a topic of high scientific interest for atmospheric and computer scientists alike. We provide data derived from the ERA5 archive that has been processed to facilitate the use in machine learning models. We propose simple and clear evaluation metrics which will enable a direct comparison between different methods. Further, we provide baseline scores from simple linear regression techniques, deep learning models, as well as purely physical forecasting models. The dataset is publicly available at \url{https://github.com/pangeo-data/WeatherBench} and the companion code is reproducible with tutorials for getting started. We hope that this dataset will accelerate research in data-driven weather forecasting.
\end{abstract}


\introduction  
\label{sec:introduction}
Deep learning, a branch of machine learning based on multi-layered artificial neural networks, has proven to be a powerful tool for a wide range of tasks, most notably image recognition and natural language processing \citep{LeCun2015}. More recently, deep learning has also been used in many fields of natural science. Much of the success of deep learning is based on the ability of neural networks to recognize patterns in high-dimensional spaces. A natural question to ask then is whether deep learning can also be used to predict future weather patterns.

Currently, weather (and climate) predictions are based on purely physical computer models, in which the governing equations, or our best approximation thereof, of the atmosphere and ocean are solved on a discrete numerical grid \citep{Bauer2015a}. Overall, this approach has been very successful. However, today's numerical weather prediction (NWP) models still have shortcoming for many important applications, for example forecasting mesoscale convective systems over Africa \citep{Vogel2018a}. Furthermore, huge amounts of computing power are required, especially for creating probabilistic forecasts which are usually limited to 50 ensemble members or less. For these reasons and the growing popularity of machine learning (ML) there has been a growing interest to improve and speed up NWP with data-driven approaches.

ML can be applied to weather prediction in many different ways. Two long-standing applications of ML are post-processing -- the correction of statistical biases in the output of physical models -- and statistical forecasting -- the prediction of variables not directly output by the physical model. Traditionally, this has been done using simple linear techniques but more recently modern machine learning approaches like random forests or neural networks have been explored \citep{Gagne2014, Taillardat2016, McGovern2017, Lagerquist2017, Rasp2018d}. Typically, these approaches target very specific variables or locations whereas the general evolution of the atmosphere is still predicted by a physical model. Another application that has recently been explored using ML is nowcasting, which describes the short range (up to 6 hours) prediction of precipitation by directly extrapolating radar observation without a physical model involved \citep{Shi2015, Shi2017, Agrawal2019, Gronquist2020}. 

Yet another direction for ML research is hybrid modeling, in which a physical model is combined with data-driven components, for example replacing heuristic cloud or radiation parameterizations \citep{Chevallier1998, Krasnopolsky2005, Rasp2018c, Brenowitz2018, Yuval2020}. The key idea behind these approaches is to only replace uncertain (e.g. clouds) or computationally expensive (e.g. line-by-line radiation) model components with machine learning emulators and leave other model components (e.g. large-scale dynamics) untouched. However such hybrid models also have drawbacks. First, the interaction between physical and machine learning components are poorly understood and can lead to unexpected instabilities and biases \citep{Brenowitz2019}. Second, they are difficult to implement from a technical perspective because one has to interface the machine learning components with complex climate model code, typically written in Fortran. 

Here we focus on purely-data driven prediction of the global atmospheric flow in the medium-range. Specifically, we select lead times of 3 and 5 days, for which the atmosphere is still reasonably deterministic but also exhibits complex nonlinear behaviour, such as baroclinic instabilities and tropical cyclogenesis. This forecast range is important from a societal point of view because it delivers crucial information for disaster preparation, for example for flooding, cold and hot spells or damaging winds \citep{Lazo2009}. Creating a good medium-range forecast requires understanding complex atmospheric dynamics and the interplay between several variables across a range of scales. This sets this challenge apart from post-processing and statistical forecasting, in which the large-scale dynamics are predicted by a physical model, and nowcasting, in which the considered evolution is univariate and short-term. In other words, this benchmark closely emulates the task performed by physical NWP models.

There are several motivations for considering a purely-data driven approach. As mentioned above current NWP is computationally expensive and, nevertheless, has low skill for certain applications. If data-driven models were able to learn a more efficient representation of the underlying dynamical and physical equations, they might enable computationally cheaper forecasts. This can be useful for many applications, for example creating very large ensembles to better estimate the probability of extreme events. It is also possible that by learning from a diverse set of data sources, data-driven models can outperform physical models in areas where the latter struggle. While in this benchmark challenge the focus is on upper-level fields of pressure and temperature -- for which physical models perform very well -- the hope is that the insights gained from this task can be leveraged for more impactful application. Further, recent research into interpretable machine learning might provide scientists with new analysis tools \citep{McGovern2019a, Toms2019}. Finally, there is the basic scientific question to what extent purely data-driven models can learn the underlying dynamics of the atmosphere.

Note also that while this benchmark is framed as a data-driven prediction challenge, the proposed framework can also be applied to post-processing using the same metrics.

In machine learning research, the data-driven prediction of future states is an active area of research with applications from language translation \citep{Sutskever2014}, over audio signals \citep{Oord2016}, to numerical simulations \citep{Morton2018}. In this context, weather forecasts are a particularly challenging task. The behavior is highly complex and non-linear, but also exhibits some recurring patterns, albeit only on local scales \citep{Hamill2006}. As such the the proposed benchmark poses interesting challenges for deep learning algorithms, e.g., to evaluate different architectures \citep{Ronneberger2015,He2015,Huang2016}, regularization methods \citep{Krogh1992,Srivastava2014,Xie2017} or optimizers \citep{Graves2013,Kingma2014}.

In the last couple of years, several studies (summarized in Section~\ref{sec:previous}) have pioneered data-driven, global, medium-range weather prediction. All of them show that there is some potential in this approach but also highlight the need for further research. In particular, we currently lack a common benchmark challenge to accelerate progress. Benchmark datasets can have a huge impact because they make different algorithms quantitatively inter-comparable and foster constructive competition, particularly in a nascent direction of research. Famous examples are the computer vision datasets MNIST \citep{LeCun1998} and ImageNet \citep{Russakovsky2015}. Further, well-curated benchmark datasets make it easier for people from different fields to work on a problem \citep{Ebert-Uphoff2017}. 

Here, we propose a benchmark problem for data-driven weather forecasting. We provide a ready-to-use dataset for download along with specific metrics to compare different approaches. In this paper, we start by reviewing the previous work done on this topic (Section~\ref{sec:previous}), describe the dataset (Section~\ref{sec:dataset}) and the evaluation metrics (Section~\ref{sec:evaluation}), and provide several baseline models (Section~\ref{sec:baselines}). Finally, we will highlight several promising directions for further research (Section~\ref{sec:discussion}) and conclude with a big picture view (Section~\ref{sec:conclusions}).

\begin{figure*}[t]
 \includegraphics[width=\linewidth]{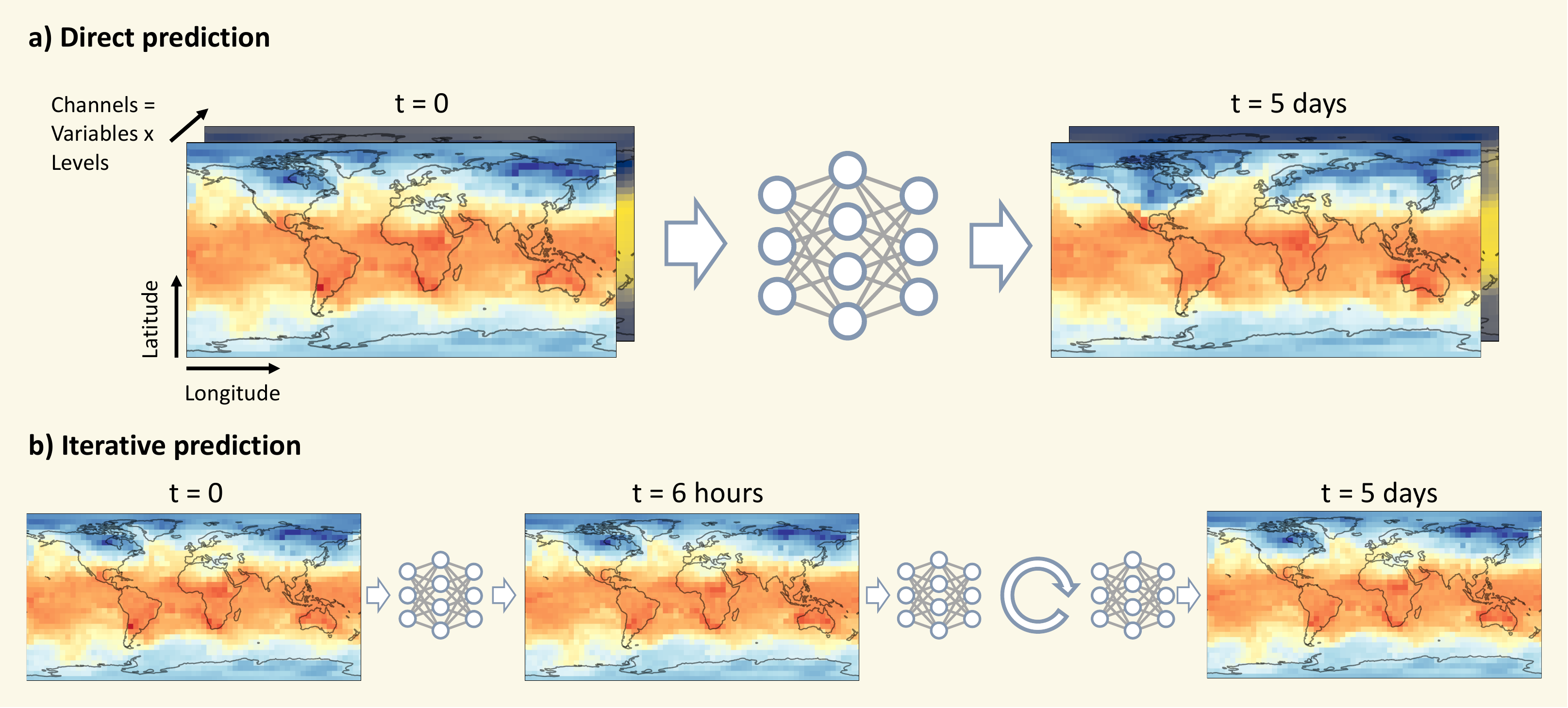}
\caption{Schematic of data-driven weather forecasting. a) Example of direct weather prediction for 5 days lead time. The input to the neural network are fields on a latitude-longitude grid. The fields can be several levels of the same variable and/or different variables. The goal is to predict the same fields some time ahead. b) Iterative forecasts are created from data-driven models trained on a shorter lead time, for example 6 hours, which are then iteratively called up to the required forecast lead time.}
\label{fig:1}
\end{figure*}

\section{Overview of previous work}
\label{sec:previous}
In this section, we briefly describe the four existing studies on predicting the large-scale atmospheric state in the medium-range with a focus on the data, methods and evaluation. 

\subsection{\cite{Dueben2018}}
In this study, the authors trained a neural network to predict 500\,hPa geopotential (Z500; see Section \ref{sec:evaluation} for details on commonly used fields), and in some experiments 2\,meter-temperature, 1 hour ahead. The training data was taken from the ERA5 archive for the time period from 2010 to 2017 and regridded to a 6 degree latitude-longitude grid. Two neural network variants were used, a fully connected neural network and a spatially localized network, similar to a convolutional neural network (CNN). After training they then created iterative forecasts up to 120\,h lead time for 10 month validation period. They compared their data-driven forecasts to an operational NWP model and the same model run at a spatial resolution comparable to the data-driven method. One interesting detail is that their networks predict the difference from one time step to the next, instead of the absolute field. To create these iterative forecasts, they use a third-order Adams-Bashford explicit time-stepping scheme. The CNN predicting only geopotential performed best but was unable to beat the low-resolution physical baseline.

\subsection{\cite{Scher2018} and \cite{Scher2019a}}
These two studies addressed the issue of data-driven weather forecasting in a simplified reality setting. Long runs of simplified General Circulation Models (GMCs) were used as ``reality''. Neural networks were trained to predict the model fields several days ahead. The neural network architecture are CNNs with an encoder-decoder setup. They take as input the instantaneous 3D model fields at one timestep, and output the same model fields at some time later. In \citet{Scher2018}, a separate network was trained for each lead-time up to 14 days. \citet{Scher2019a} trained only on 1-day forecasts, and constructed longer forecasts iteratively. Interestingly, networks trained to directly predict a certain forecast time, e.g. 5 days, outperformed iterative networks. The forecasts were evaluated using the root mean squared error and the anomaly correlation coefficient of Z500 and 800\,hPa temperature. \citet{Scher2018} used a highly simplified GCM without hydrological cycle, and achieved very high predictive skill. Additionally, they were able to create stable "climate" runs (long series of consecutive forecasts) with the network. \citet{Scher2019a} used several more realistic and complex GCMs. The data-driven model achieved relatively good short-term forecast skill, but was unable to generate stable and realistic ``climate'' runs. In terms of neural-network architectures they showed that architectures tuned on simplified GCMs also work on more complex GCMs, and that the same architecture also has some prediction skill on single-level reanalysis data.

\subsection{\cite{Weyn2019}}
In this study, reanalysis-derived Z500 and 700-300\,hPa thickness at 6-hourly time steps are predicted with deep CNNs. The data are from the Climate Forecast System (CFS) Reanalysis from 1979--2010 with 2.5-degree horizontal resolution and cropped to the northern hemisphere. The authors used similar encoder-decoder convolutional networks as those used by \cite{Scher2018} and \cite{Scher2019a} but also experimented with adding a convolution long short-term memory \citep[LSTM;][]{Hochreiter1997} hidden layer. As in \cite{Scher2019a}, forecasts are generated iteratively by feeding the model's outputs back in as inputs. The authors found that using two input time steps, 6\,h apart, and predicting two output time steps, performed better than using a single step. Their best CNN forecast outperforms a  climatology benchmark at up to 120\,h lead time, and appears to correctly asymptote towards persistence forecasts at longer lead times up to 14 days. 

These three approaches outline promising first steps towards data-driven forecasting. The differences of the proposed methods already highlight the importance of a common benchmark case to compare prediction skill.

\section{Dataset}
\label{sec:dataset}
For the proposed benchmark, we use the ERA5 reanalysis dataset \citep{Hersbach2020} for training and testing. Reanalysis datasets provide the best guess of the atmospheric state at any point in time by combining a forecast model with the available observations. 
The raw data is available hourly for 40 years from 1979 to 2018 on a 0.25\textdegree latitude-longitude grid (721$\times$1440 grid points) with 37 vertical levels. 

Since this raw dataset is very large (a single vertical level for the entire time period amounts to almost 700GB of data), we regrid the data to lower resolutions. This is also a more realistic use case, since very high resolutions are still hard to handle for deep learning models because of GPU memory constraints and I/O speed. In particular, we chose 5.625\textdegree \ (32$\times$64 grid points),  2.8125\textdegree \ (64$\times$128 grid points) and 1.40525\textdegree \ (128$\times$256 grid points) resolution for our data. The regridding was done with the xesmf Python package \citep{Zhuang2019} using a bilinear interpolation. Powers of two for the grid are used since this is common for many deep learning architectures where image sizes are halved in the algorithm. Further, for 3D fields we selected 13 vertical levels: 50, 100, 150, 200, 250, 300, 400, 500, 600, 700, 850, 925, 1000\,hPa. Note that it is common to use pressure in hecto-Pascals as a vertical coordinate instead of physical height. The pressure at sea level is approximately 1000\,hPa and decreases roughly exponentially with height. 850\,hPa is at around 1.5\,km height. 500\,hPa is at around 5.5\,km height. If the surface pressure is smaller than a given pressure level, for example at high altitudes, the pressure-level values are interpolated. The selected pressure levels contain the seven pressure levels that are commonly used for 3D output by the climate models in the Coupled Model Intercomparison Project Phase 6 \citep[CMIP6,][]{Eyring2016} which could be useful for pretraining. One regridded historical climate run is also available from the data repository with a template workflow for downloading further CMIP data on the Github repository.

The processed data (see Table~\ref{tab:1}) are available at \url{https://mediatum.ub.tum.de/1524895} \citep{Rasp2020Data}. The data are split into yearly NetCDF files for each variable and resolution, packed in a zip file. The entire dataset at 5.625\textdegree \ resolution has a size of 191GB. Individual variables amount to around 25GB three-dimensional and 2GB for two-dimensional fields. File sizes for 2.8125\textdegree \ and 1.40525\textdegree \  resolutions are a factor 4 and 16 times larger. Data processing was organized using Snakemake \citep{Koster2012}. For further instructions on data downloading visit the Github page\footnote{\url{https://github.com/pangeo-data/WeatherBench}}. The available variables were chosen based on meteorological consideration. Geopotential, temperature, humidity and wind are prognostic state variables in most physical NWP and climate models. Geopotential at a certain pressure level $p$, typically denoted as $\Phi$ with units of m$^2$s$^{-2}$, defined as 
\begin{equation}
    \Phi = \int_0^{z \text{ at } p} g\,dz'
\end{equation}
where $z$ describes height in meters and $g=9.81$ m s$^{-2}$ is the gravitational acceleration. Horizontal relative vorticity, defined as $\partial v / \partial x - \partial u / \partial y$, describes the rotation of air at a given point in space. Potential vorticity \citep{Hoskins1985, Holton2004} is a commonly used quantity in synoptic meteorology which combines the rotation (vorticity) and vertical temperature gradient of the atmosphere. It is defined as $PV = \rho^{-1} \zeta_a \cdot \nabla \theta$, where $\rho$ is the density, $\zeta_a$ is the absolute vorticity (relative plus the Earth's rotation) and $\theta$ is the potential temperature. In addition to the three-dimensional fields, we also include several two-dimensional fields: 2\,meter-temperature is often used as an impact variable because of its relevance for human activities and is directly affected by the diurnal solar cycle; 10\,meter-wind is also an important impact-related forecast variable, for example for wind energy; similarly, total cloud cover is an essential variable for solar energy forecasting. We also included precipitation but urge caution since precipitation in reanalysis datasets often shows large deviation from observations \citep[e.g.][]{Betts2019, Xu2019}. Finally, we added the top-of-atmosphere incoming solar radiation as it could be a useful input variable to encode the diurnal cycle. Further, there are several potentially important time-invariant fields, which are contained in the constants file. The first three variables enclose information about the surface: the land-sea mask is a binary field with ones for land points; the soil type consists of seven different soil categories\footnote{Coarse = 1, Medium	= 2, Medium fine = 3, Fine = 4, Very fine = 5, Organic = 6, Tropical organic = 7, see \url{https://apps.ecmwf.int/codes/grib/param-db?id=43}}; orography is simply the surface height. In addition, we included two-dimensional fields with the latitude and longitude values at each point. Particularly the latitude values could become important for the network to learn latitude-specific information such as the grid structure or the Coriolis effect (see Section~\ref{sec:discussion}). The Github code repository includes all scripts for downloading and processing of the data. This enables users to download additional variables or regrid the data to a different resolution.

\begin{table}[t]
\caption{List of variables contained in the benchmark dataset.}
\label{tab:1}
\begin{tabular}{l|l|l|l|l}
\textbf{Long name}         & \textbf{Short name}              & \textbf{Description}                                         & \textbf{Unit}      & \textbf{Levels}                         \\ \hline \hline
geopotential & z               & Proportional to the height of a pressure level      & [m$^2$s$^{-2}$] & 13 levels    \\ \hline
temperature & t                &      Temperature                                               & [K]       & 13 levels   \\ \hline
specific\_humidity & q          & Mixing ratio of water vapor                        & [kg kg$^{-1}$] & 13 levels                       \\ \hline
relative\_humidity & r         & Humidity relative to saturation                     & [\%]       & 13 levels                       \\ \hline
u\_component\_of\_wind & u                 & Wind in x/longitude-direction                       & [m s$^{-1}$]     & 13 levels             \\ \hline
v\_component\_of\_wind & v            & Wind in y/latitude direction                        & [m s$^{-1}$]     & 13 levels                \\ \hline
vorticity & vo      & Relative horizontal vorticity                                      & [1 s$^{-1}$]     & 13 levels                       \\ \hline
potential\_vorticity & pv      & Potential vorticity                                      & [K m$^2$ kg$^{-1}$ s$^{-1}$]     & 13 levels                       \\ \hline
2m\_temperature & t2m      & Temperature at 2 m height above surface             & [K]       & Single level                    \\ \hline
10m\_u\_component\_of\_wind & u10      & Wind in x/longitude-direction at 10 m height             & [m s$^{-1}$]       & Single level                    \\ \hline
10m\_v\_component\_of\_wind & v10      & Wind in y/latitude-direction at 10 m height             & [m s$^{-1}$]       & Single level                    \\ \hline
total\_cloud\_cover &   tcc    & Fractional cloud cover             &     (0--1)   & Single level                    \\ \hline
total\_precipitation & tp      & Hourly precipitation             &   [m]     & Single level                    \\ \hline
toa\_incident\_solar\_radiation & tisr & Accumulated hourly incident solar radiation              & [J m$^{-2}$]        & Single level       \\ \hline \hline
\textbf{constants} &               & \textit{File containing time-invariant fields} &          &       \\ 
land\_binary\_mask &        lsm       & Land-sea binary mask &        (0/1)   & Single level       \\
soil\_type & slt & Soil-type categories & see text & Single level \\
orography & orography & Height of surface & [m] & Single level \\
latitude & lat2d & 2D field with latitude at every grid point & [\textdegree] & Single level \\
longitude & lon2d & 2D field with longitude at every grid point & [\textdegree] & Single level \\
\end{tabular}
\belowtable{} 
\end{table}

\section{Evaluation}
\label{sec:evaluation}
Evaluation is done for the years 2017 and 2018. To make sure no overlap exists between the training and test dataset, the first test date is 1 January 2017 00UTC plus forecast time (i.e. for a three day forecast the first test date would be 4 January 2017 00UTC) while the last training target is 31 December 2016 23UTC. Further, the evaluation presented here is done on 5.625\textdegree \ resolution\footnote{The evaluation of all baselines in this paper are done in this Jupyter notebook: \url{https://github.com/pangeo-data/WeatherBench/blob/master/notebooks/4-evaluation.ipynb}}.  This means that predictions at higher resolutions have to be downscaled to the evaluation resolution. We also evaluated some baselines at higher resolutions and found that the scores were almost identical with differences smaller than 1\%. Therefore we are reassured that little information is lost by evaluating at a coarser resolution.

A note on validation and testing: In machine learning it is good practice to split the data into three parts: the training, validation and test sets. The training dataset is used to actually fit the model. The validation dataset is used during experimentation to check the model performance on data not seen during training. However, there is the danger that through continued tuning of hyperparameters one unwillingly overfits to the validation dataset. Therefore it is advisable to keep a third, testing, dataset for final evaluations of model performance. 
For this benchmark this final evaluation is done for the years 2017 and 2018. Therefore, we strongly encourage users of this dataset to pick a period from 1979 to 2016 for validation of their models for hyperparameter tuning. Because meteorological fields are highly correlated in time, it is advisable to choose a longer contiguous chunk of data for validation instead of a completely random split. Here we chose the year 2016 for validation.

We chose 500\,hPa geopotential and 850\,hPa temperature as primary verification fields. Geopotential at 500\,hPa pressure, often abbreviated as Z500, is a commonly used variable that encodes the synoptic-scale pressure distribution. It is the standard verification variable for most medium-range NWP models. Note that \textit{geopotential height}, also commonly used, is defined as $\Phi / g$ with units of meters. We picked 850\,hPa temperature as our secondary verification field because temperature is a more impact-related variable. 850\,hPa is usually above the planetary boundary layer and therefore not affected by diurnal variations but provides information about broader temperature trends, including cold spells and heat waves. In addition we also provide some baseline scores for total 6-hourly accumulated precipitation (TP; but noting the dubious quality mentioned above) and 2-meter temperature (T2M). However, they will not be further discussed here.

We chose the root mean squared error (RMSE) as our primary metric because it is easy to compute and mirrors the loss used for most ML applications. We define the RMSE as the mean latitude-weighted RMSE over all forecasts:
\begin{equation}
\label{eq:rmse}
    \text{RMSE} = \frac{1}{N_{\text{forecasts}}} \sum_i^{{N_{\text{forecasts}}}} \sqrt{\frac{1}{N_{\text{lat}} N_{\text{lon}}} \sum_j^{N_{\text{lat}}} \sum_k^{N_{\text{lon}}} L(j) (f_{i, j, k} - t_{i, j, k})^2} 
\end{equation}
where $f$ is the model forecast and $t$ is the ERA5 truth. $L(j)$ is the latitude weighting factor for the latitude at the $j$th latitude index:
\begin{equation}
    L(j) = \frac{\cos( \text{lat}(j))}{\frac{1}{N_{\text{lat}}} \sum_j^{N_{\text{lat}}} \cos( \text{lat}(j)) }
\end{equation}

In addition, we also evaluate the baselines using the latitude weighted anomaly correlation coefficient \citep[ACC; see Section 7.6.4 of][]{Wilks2006} and the mean absolute error (MAE). The tables and figures can be found in the Appendix~\ref{sec:app}. For smooth fields like Z500 and T850 the qualitative differences between the metrics are small. For intermittent fields like precipitation the choice of metric matters a lot more.

\section{Baselines}
\label{sec:baselines}
To evaluate the skill of a forecasting model it is important to have baselines to compare to. In this section, we compute scores for several baselines. The results are summarized in Fig.~\ref{fig:rmse} and Table~\ref{tab:scores}.

\begin{figure*}[t]
 \includegraphics[width=\linewidth]{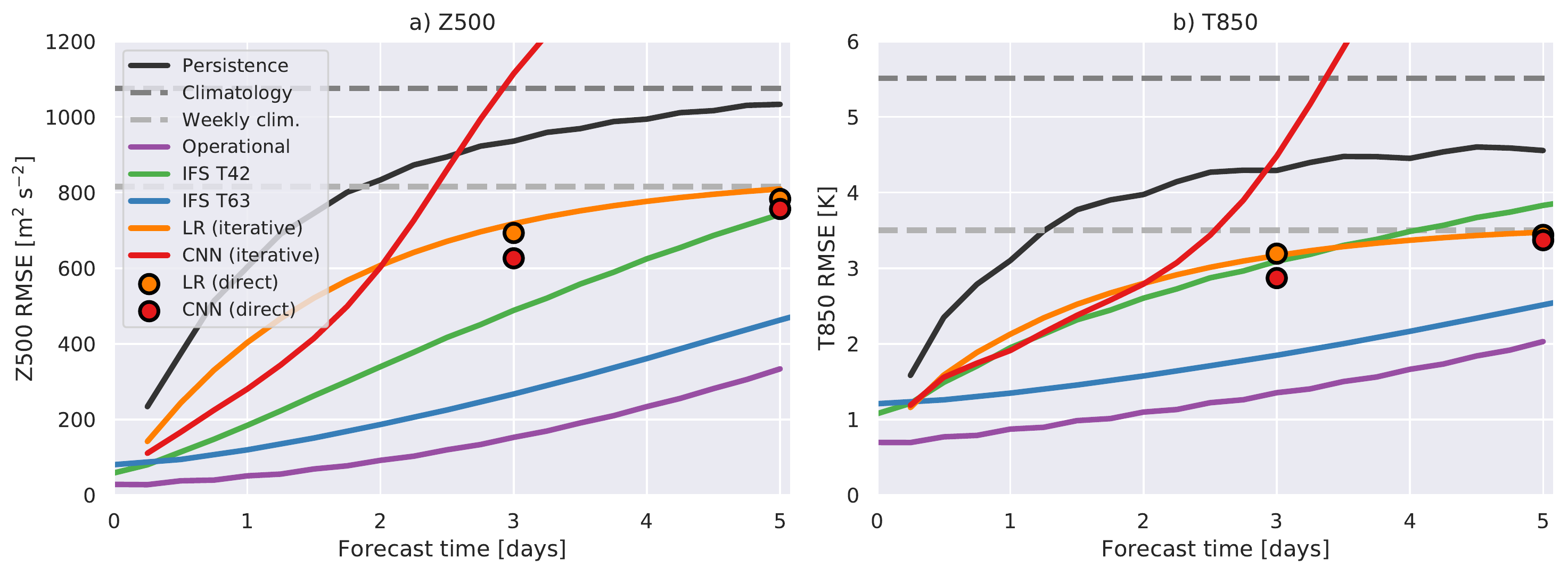}
\caption{RMSE of a) 500\,hPa geopotential and b) 850\,hPa temperature for different baselines at 5.625\textdegree \ resolution. Solid lines for linear regression and CNN indicate iterative forecasts, while dots represent direct forecasts for 3 and 5 days lead time.}
\label{fig:rmse}
\end{figure*}

\begin{table}[t]
\caption{Baseline RMSE for 3 and 5 days forecast time at 5.625\textdegree \ resolution. Best machine learning baseline and physical model are highlighted. TP is 6 hourly accumulated precipitation.}
\begin{tabular}{l||l|l|l|l}
 \multicolumn{4}{c}{\hspace{180pt} RMSE (3 days / 5 days)} \\
\textbf{Baseline} & \textbf{Z500 [m$^2$ s$^{-2}$]} & \textbf{T850 [K]} & \textbf{T2M [K]} & \textbf{TP [mm]}\\ \hline \hline
Persistence & 936 / 1033 & 4.23 / 4.56 & 3.00 / 3.27 & 3.23 / 3.24 \\ \hline
Climatology & 1075 & 5.51  & 6.07 & 2.36 \\ \hline
Weekly climatology & 816 & 3.50  & 3.19 & \textbf{2.32} \\ \hline
Linear regression (direct) & 693 / 783 & 3.19 / 3.44 & 2.39 / 2.60 & 2.37 / 2.37 \\ \hline
Linear regression (iterative) & 718 / 810 & 3.17 / 3.48  & & \\ \hline
CNN (direct) & \textbf{626} / \textbf{757} & \textbf{2.87} / \textbf{3.37}  & & \\ \hline
CNN (iterative) & 1114 / 1559 & 4.48 / 9.69  & & \\ \hline
IFS T42 & 489 / 743 & 3.09 / 3.83  & 3.21 / 3.69 & \\ \hline
IFS T63 & 268 / 463 & 1.85 / 2.52  & 2.04 / 2.44 & \\ \hline
Operational IFS & \textbf{154} / \textbf{334} & \textbf{1.36} / \textbf{2.03} & \textbf{1.35} / \textbf{1.77} & 2.36 / 2.59
\end{tabular}
\label{tab:scores}
\end{table}

\subsection{Persistence and Climatology}
The two simplest possible forecasts are a) a persistence forecast in which the fields at initialization time are used as forecasts ("tomorrow's weather is today's weather"), and b) a climatological forecast. For the climatological forecast, two different climatologies were computed from the training dataset (1979--2016): first, a single mean over all times in the training dataset and, second, a mean computed for each of the 52 calendar weeks. The weekly climatology is significantly better, approximately matching the persistence forecast between 1 and 2 days, since it takes into account the seasonal cycle. This means that to be useful, a forecast system needs to beat the weekly climatology and the persistence forecast. 

\subsection{Operational NWP model}
The gold standard of medium-range NWP is the operational IFS (Integrated Forecast System) model of the European Center for Medium-range Weather Forecasting (ECMWF)\footnote{\url{https://www.ecmwf.int/en/forecasts/documentation-and-support}}. We downloaded the forecasts for 2017 and 2018 from the THORPEX Interactive Grand Global Ensemble \citep[TIGGE;][]{Bougeault2010} archive\footnote{The TIGGE data for total precipitation and 2m temperature was damaged for 2017. For this reason the TIGGE evaluation for these variables is only done using the 2018 data.}, which contains the operational forecasts, initialized at 00 and 12 UTC regridded to a 0.5\textdegree \ by 0.5\textdegree \ grid, which we further regridded to 5.625\textdegree. Note that the forecast error starts above zero because the operational IFS is initialized from a different analysis. Operational forecasting is computationally very expensive. The current IFS deterministic forecast is computed on a cluster with 11,664 cores. One 10 day forecast at 10\,km resolution takes around 1 hour of real time to compute. 

\subsection{Physical NWP model run at coarser resolution}
To provide physical baselines more in line with the computational resources of a data-driven model, we ran the IFS model at two coarser horizontal resolutions, T42 (approximately 2.8\textdegree \ or 310\,km resolution at the equator \citep{NCAR2020}) with 62 vertical levels and T63 (approximately 1.9\textdegree \ or 210\,km) with 137 vertical levels. The T42 run was initialized from ERA5 whereas the T63 run was initialized from the operational analysis. The gap in skill at $t=0$ is caused by the conversion to spherical coordinates at coarse resolutions. For Z500 the skill for these two runs lies in-between the operational IFS and the machine learning baselines. For T850, the T42 run is significantly worse. The likely reason for this is that temperature close to the ground is much more affected by the resolution and representation of topography within the model. Further, the model was not specifically tuned for these resolutions. Computationally, a single forecast takes 270 seconds for the T42 model and 503 seconds for the T64 model on a single XC40 node with 36 cores. Since the computational costs and resolutions of these runs are much closer to those of a data-driven method, beating those baselines should be a realistic target. note, however, that the model was not tuned to run at such coarse resolutions. 

\subsection{Linear regression}
As a first purely data-driven baseline we fit a simple linear regression model. For the direct predictions a separate model was trained for each of the four variables. For this purpose the 2D fields were flattened from 32$\times$64 $\to$ 2048. This was done for 3\,d and 5\,d forecast time. In addition an iterative model for Z500 and T850 was trained. Here we use a single linear regression to predict 6 hours ahead where the two fields are concatenated (2$\times$32$\times$64 $\to$ 4096). The advantage of iterative forecasts is that a single model is able to make predictions for any forecast time rather than having to train several models. For iterative forecasts the model takes its previous output as input for the next step. To create a 5 day iterative forecast the model trained to predict 6 hour forecasts is called 20 times. For this model, the iterative forecast performs just as well as the direct forecast due to its linear nature. At 5 days, the linear regression forecast is about as good as the weekly climatology.

\subsection{Simple convolutional neural network}
As our deep learning baseline we chose a simple fully-convolutional neural network. CNNs are the natural choice for spatial data since they exploit translational invariances in images/fields. Here we train a CNN with 5 layers. Each hidden layer has 64 channels with a convolutional kernel of size 5 and ELU activations \citep{Clevert2015}. The input and output layers have two channels, representing Z500 and T850. The model was trained using the Adam optimizer \citep{Kingma2014} and a mean squared error loss function. The total number of trainable parameters is 313,858. We implemented periodic convolutions in the longitude direction but not the latitude direction. The implementation can be found in the Github repository. The direct CNN forecasts beat the linear regression forecasts for 3 and 5 days forecast time. However, at 5 days these forecasts are only marginally better than the weekly climatology (see Table~\ref{tab:scores}). This baseline, however, is to be seen simply as a starting point for more sophisticated data driven methods. The iterative CNN forecast, which equivalently to the linear regression iterative forecast was created by chaining together 6 hourly predictions, performs well up to around 1.5 days but then the network's errors grow quickly and diverge. This confirms the findings of \cite{Scher2019} whose experiments showed that training with longer lead time yields better results than chaining together short-term forecasts. However, the poor skill of the iterative forecast could easily be a result of using an overly simplistic network architecture. The iterative forecasts of \cite{Weyn2019}, who employ a more complex network structure, show stable long term performance up to two weeks with realistic statistics.

\subsection{Example forecasts}
\begin{figure*}[t]
 \includegraphics[width=\linewidth]{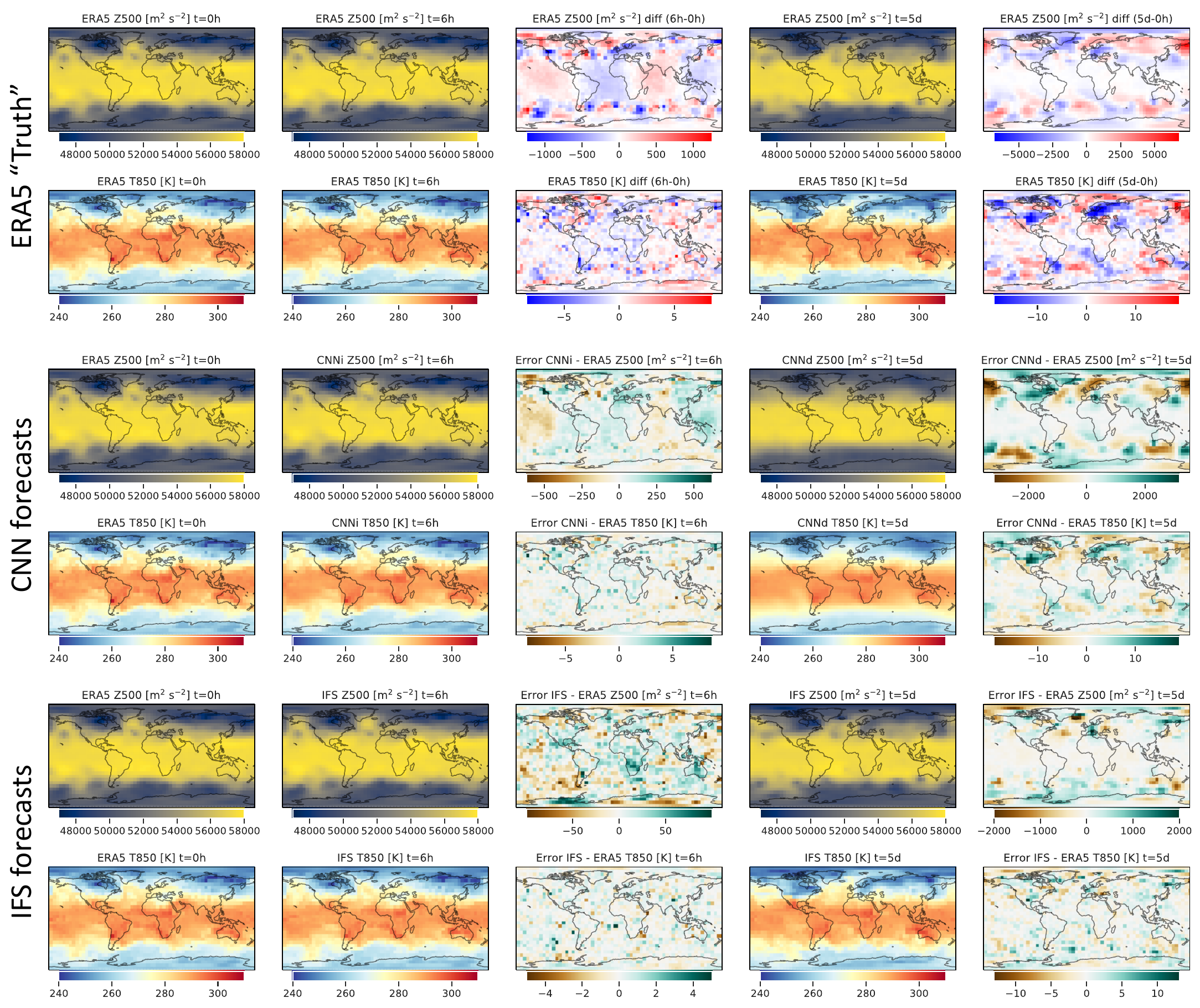}
\caption{Example fields for 2017-01-01 00UTC initialization time. The top two rows show the ERA5 "truth" fields for geopotential (Z500) and temperature (T850) at initialization time (t=0h) and for 6h and 5d forecast time. In addition, the difference between the forecast times and the initialization time is shown. The third and fourth rows show the forecasts from the CNN model. Rows five and six show the IFS operational model. For the CNN forecasts the first column is identical to the ERA5 truth. We selected the 6h iterative CNN model for the 6h forecast but the 5d direct CNN model for the 5 day forecast. For the IFS the initial states (t=0h) differ slightly albeit not visibly. In addition to the forecast fields the error relative to the ERA5 "truth" is shown in the third and fifth columns. Please note that the colorbars for the difference fields change.}
\label{fig:examples}
\end{figure*}

To further illustrate the prediction task, Fig.~\ref{fig:examples} shows example geopotential and temperature fields. The ERA5 temporal differences show several interesting features. First, the geopotential fields and differences are much smoother compared to the temperature fields. The differences in both fields are also much smaller in the tropics compared to the extratropics where propagating fronts can cause rapid temperature changes. An interesting feature is detectable in the 6h Z500 difference field in the tropics. These alternating patterns are signatures of atmospheric tides. 

The CNN forecasts for 6h lead time are not able to capture these wave-like patterns which hints at a failure to capture the basic physics of the atmosphere. For 5 days forecast time the CNN model predicts unrealistically smooth fields. This is likely a result of two factors: first, the two input fields used in this baseline CNN contain insufficient information to create a skillful 5 day forecast; and second, at 5 days the atmosphere already shows some chaotic behavior which causes a model trained with a simple RMSE loss to predict smooth fields (see Section~\ref{sec:discussion}). The IFS operational forecast has much smaller errors than the CNN forecast. It is able to capture the propagation of tropical waves. Its main errors appear at 5 days in the mid-latitudes where extratropical cycles are in slightly wrong positions.

\section{Discussion}
\label{sec:discussion}
\subsection{Weather-specific challenges}
From a ML perspective, state-to-state weather prediction is similar to image-to-image translation. For this sort of problem many deep learning techniques have been developed in recent years \citep{Kaji2019}. However, forecasting weather differs in some important ways from typical image-to-image applications and raises several open questions.

First, the atmosphere is three-dimensional. So far, this aspect has not been taken into account. In the networks of \cite{Scher2019}, for example, the different levels have been treated as separate channels of the CNN. However, simply using a three-dimensional CNN might not work either because atmospheric dynamics and grid spacings change in the vertical, thereby violating the assumption of translation invariance which underlies the effectiveness of CNNs. This directly leads to the next challenge: On a regular latitude-longitude grid, the dynamics also change with latitude because towards the poles the grid cells become increasingly stretched. This is in addition to the Coriolis effect, the deflection of wind caused by the rotation of Earth, which also depends on latitude. A possible solution in the horizontal could be to use spherical convolutions \citep{Cohen2018,Perraudin2019,Jiang2019} or to feed in latitude information to the network. 

Another potential issue is the limited amount of training data available. 40 years of hourly data amounts to around 350,000 samples. However, the samples are correlated in time. If one assumes that a new weather situation occurs every day, then the number of samples is reduced to around 15,000. Without empirical evidence it is hard to estimate whether this number is sufficient to train complex networks without overfitting. Should overfitting be a problem, one could try transfer learning. In transfer learning, the network is pretrained on a similar task or dataset, for example, climate model simulations, and then finetuned on the actual data. This is common practice in computer vision and has been successfully applied to seasonal ENSO forecasting \citep{Ham2019}. Another common method to prevent overfitting is data augmentation, which in traditional computer vision is done by e.g. randomly rotating or flipping the image. However, many of the traditional data augmentation techniques are questionable for physical fields. Random rotations, for example, will likely not work for this dataset since the x and y directions are physically distinct. Thus, finding good data augmentation techniques for physical fields is an outstanding problem. Using ensemble analyses and forecasts could provide more diversity in the training dataset. 

Finally, there are technical challenges. Data for a single variable with ten levels at 5.625\textdegree \ resolution take up around 30\,GB of data. For a network with several variables or even at higher resolution, the data might not fit into CPU RAM any more and data loading could become a bottleneck. For image files, efficient data loaders have been created\footnote{See e.g. \url{https://keras.io/preprocessing/image/} or \url{https://pytorch.org/tutorials/beginner/data_loading_tutorial.html}. One promising but so far unexplored option is to use Tensorflow's TFRecords (\url{https://www.tensorflow.org/tutorials/load_data/tfrecord})}. For netCDF files, however, so far no efficient solution exists to our knowledge. Further, one can assume that to create a competitive data-driven NWP model, high resolutions have to be used, for which GPU RAM quickly becomes a limitation. This suggests that multi-GPU training might be necessary to scale up this approach (potentially similar to the technical achievement of \cite{Kurth2018}).

\subsection{Probabilistic forecasts and extremes}
One important aspect that is not currently addressed by this benchmark is probabilistic forecasting. Because of the chaotic evolution of the atmosphere, it is very important to also have an estimate of the uncertainty of a forecast. In physical NWP this is done by running several forecasts, called an ensemble, from slightly different initial conditions and potentially with different or stochastic model physics \citep{Palmer2019b}. From this Monte Carlo forecast one can then estimate a probability distribution. A different approach, which is often taken in statistical post-processing of NWP forecasts, is to directly estimate a parametric distribution \citep[e.g.][]{Gneiting2005, Rasp2018d}. For a probabilistic forecast to be reliable the forecast uncertainty has to be an accurate indicator of the error. A good first order approximation for this is the spread (ensemble standard deviation) to error (RMSE) ratio which should be one \citep{Leutbecher2008}. A more stringent test is to use a proper probabilistic scoring rule, for example the continuous ranked probability score (CRPS) \citep{Gneiting2007, Gneiting2014a}. For deterministic forecast the CRPS reduces to the mean absolute error. Extending this benchmark to probabilistic forecasting simply requires computing a probabilistic score. How to produce probabilistic data-driven forecasts is a very interesting research question in its own right. We encourage users of this benchmark to explore this dimension.

A related issue is the question of extreme weather situations, for example heat waves. These events are, by definition, rare, which means that they will contribute little to regular verification metrics like the RMSE. However, for society these events are highly important. For this reason, it would make sense to evaluate extreme situations separately. But defining extremes is ambiguous which is why there is no standard metric for evaluating extremes. The goal of this benchmark is to provide a simple, clear problem. Therefore, we decided to omit extremes for now but users are encouraged to chose their own verification of extremes.

\subsection{Climate simulations}
Another aspect that is untouched by the benchmark challenge proposed here is climate prediction. Even though weather and climate deal with the same underlying physical system, they pose different forecasting challenges. In weather forecasting the goal is to predict the state of the atmosphere at a specific time into the future. This is only possible up to the prediction horizon of the atmosphere, which is thought to be at roughly two weeks. Climate models, on the other hand, are evaluated by comparing long-term statistics to observations, for example the mean surface temperature \citep{Stocker2013}. \cite{Scher2019} created iterative climate time scale runs with their data-driven models and compared first and second-order statistics. They found that the model sometimes produced a stable climate but with significant biases and a poor seasonal cycle. This indicates that, so far, iterative data-driven models have been unable to produce physically reasonable long-term predictions. This remains a key challenge for future research. While not specifically included in this benchmark, a good test for climate simulations is to look at long term mean statistics and the seasonal cycle as done in Figs. 6 and 7 of \cite{Scher2019}. 

Climate change simulations represent another step up in complexity. To start with, external greenhouse gas forcing would have to be included. Further, future climates will produce atmospheric states that lie outside of the historical manifold of states. Plain neural networks are very bad at extrapolating to climates beyond what they have seen in the training dataset \citep{Rasp2018c}. For this reason, climate change simulations with current data-driven methods are likely not a good idea. However, research into physical machine learning is ongoing and might offer new opportunities in the near future \citep[e.g.][]{Bar-Sinai2019a, Beucler2019}.

\subsection{Promising research directions}
There is a wide variety of promising research directions for data-driven weather forecasting. The most obvious direction is to increase the amount of data used for training and the complexity of the network architecture. This dataset provides a, so far, unexploited volume and diversity of data for training. It is up to future research to find out exactly which combination of variables will turn out to be useful. Further, this dataset offers a four times higher horizontal resolution than all previous studies. The hope is that this data will enable researcher to train more complex models than have previously been used. 

With regards to model architecture, there is a huge variety of network architectures that can be explored. U-Nets \citep{Ronneberger2015} have been used extensively for image segmentation tasks that require computations across several spatial scales. Resnets \citep{He2015} are currently the state of the art for image classification and their residual nature could be a good fit for state-to-state forecasting tasks. For synthesis tasks, generative adverserial networks (GANs) \citep{Goodfellow2014} were shown to be particularly powerful for creating realistic natural images and fluid flows \citep{Xie2018}. This might be attractive since minimizing a mean loss, such as the MSE, for random or stochastic data leads to unrealistically smooth predictions as seen in Fig.~\ref{fig:examples}. Conditional GANs \citep{Mirza2014,Isola2016} could potentially alleviate this issue but it is still unclear to what extent GAN predictions are able to recover the multi-variate distribution of the training samples.

\codedataavailability{The dataset is available at \url{https://mediatum.ub.tum.de/1524895} \citep{Rasp2020Data}. Code, instructions for dowloading the data and evaluating forecasts can be found at \url{https://github.com/pangeo-data/WeatherBench}.}

\conclusions  
\label{sec:conclusions}
In this paper a benchmark dataset for data-driven weather forecasting is presented. It focuses on global medium-range (roughly 2 days to 2 weeks) prediction. With the rise of deep learning in physics, weather prediction is a challenging and interesting target because of the large overlap with traditional deep learning tasks \citep{Reichstein2019}. While first attempts have been made in this direction, as discussed in Section~\ref{sec:previous}, the field currently lacks a common dataset which enables the inter-comparison of different methods. We hope that this benchmark can provide a foundation for accelerated research in this area. Loosely following \cite{Ebert-Uphoff2017}, the key features of this benchmark are:
\begin{itemize}
    \item \textbf{Scientific impact:} Numerical weather forecasting impacts many aspects of society. Currently, NWP model run on massive super-computers at very high computational cost. Building a capable data-driven model would be beneficial in many ways (see Section~\ref{sec:introduction}). In addition, there is the open, and highly debated, question whether fully data-driven methods are able to learn a good representation of atmospheric physics.
    \item \textbf{Challenge for data science:} While global weather prediction is conceptually similar an image-to-image task, and therefore allows for the application of many state-of-the-art deep learning techniques, there are some unique challenges to this problem: the three-dimensional, anisotropic nature of the atmosphere; non-uniform-grids; potentially limited amounts of training data and the technical challenge of handling large data volumes.
    \item \textbf{Clear metric for success:} We defined a single metric (RMSE) for two fields (500\,hPa geopotential and 850\,hPa temperature). These scores provide a simple measure of success for data-driven, medium-range forecast model.
    \item \textbf{Quick start:} The code repository contains a quick-start Jupyter notebook for reading the data, training a neural network and evaluating the predictions against the target data. In addition, the repository contains many functions which are likely to be used frequently, for example an implementation of periodic convolutions in Keras.
    \item \textbf{Reproducibility and citability:} All baselines and results form this paper are fully reproducible from the code repository. Further, the baseline predictions are all saved in the data repository. The data has been assigned a permanent DOI.
    \item \textbf{Communication platform:} We will use the Github code repository as an evolving hub for this project. We encourage users of this dataset to start by forking the repository and eventually merge code that might be useful for others back into the main branch. The main platform for communication, e.g. asking questions, about this project will be Github issues.
\end{itemize}

We hope that this benchmark will foster collaboration between atmospheric and data scientists in the ways we imagined and beyond.







\clearpage
\appendix
\section{Additional metrics}    
\label{sec:app}

The anomaly correlation coefficient (ACC) is defined as 
\begin{equation}
    \text{ACC} = \frac{\sum_{i, j, k} L(j)  f'_{i, j, k}  t'_{i, j, k}}{\sqrt{\sum_{i, j, k} L(j)  f'^{2}_{i, j, k} \sum_{i, j, k} L(j)  t'^{2}_{i, j, k}}}
\end{equation}
where the prime $'$ denotes the difference to the climatology. Here the climatology is defined as $\mathrm{climatology}_{j, k} = \frac{1}{N_{time}} \sum t_{j, k}$. The mean absolute error is defined just like the MSE (Eq.~\ref{eq:rmse}) but with the absolute instead of the squared difference.

\appendixfigures
\begin{figure*}[th!]
 \includegraphics[width=\linewidth]{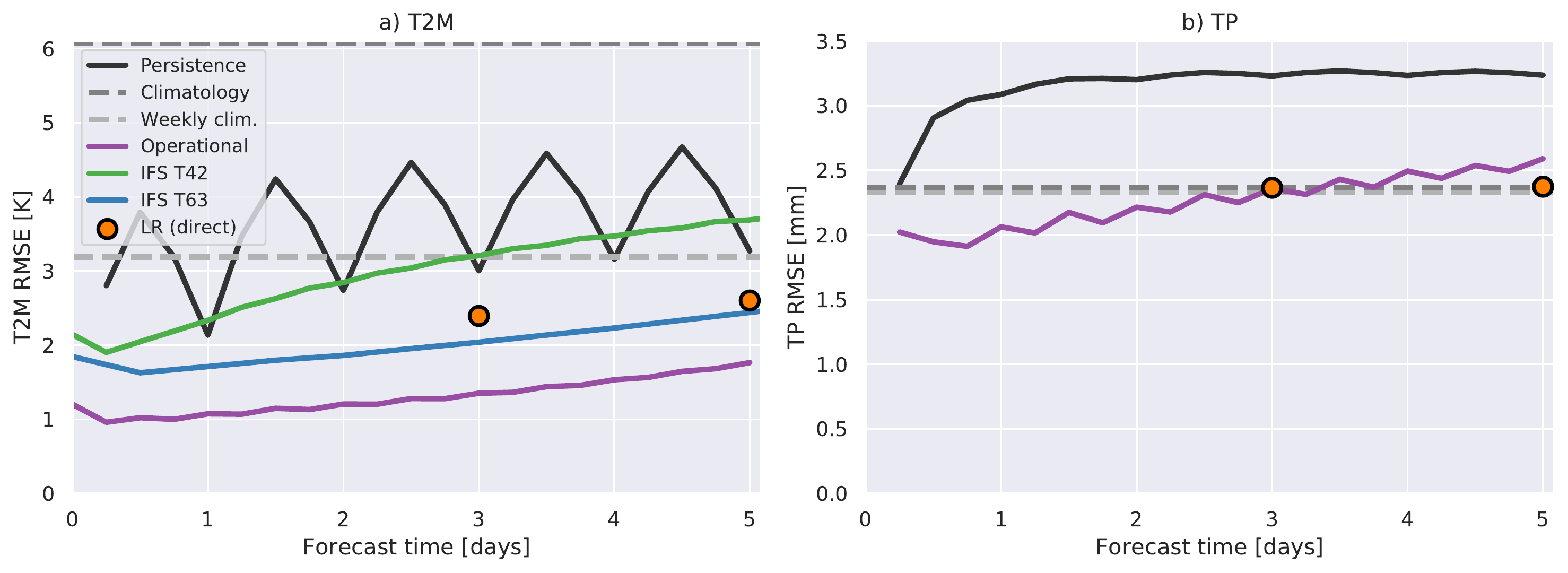}
\caption{RMSE of a) 2-meter temperature and b) 6-hourly accumulated precipitation for different baselines at 5.625\textdegree \ resolution.}
\label{fig:rmse2}
\end{figure*}
\clearpage

\begin{figure*}[th!]
 \includegraphics[width=\linewidth]{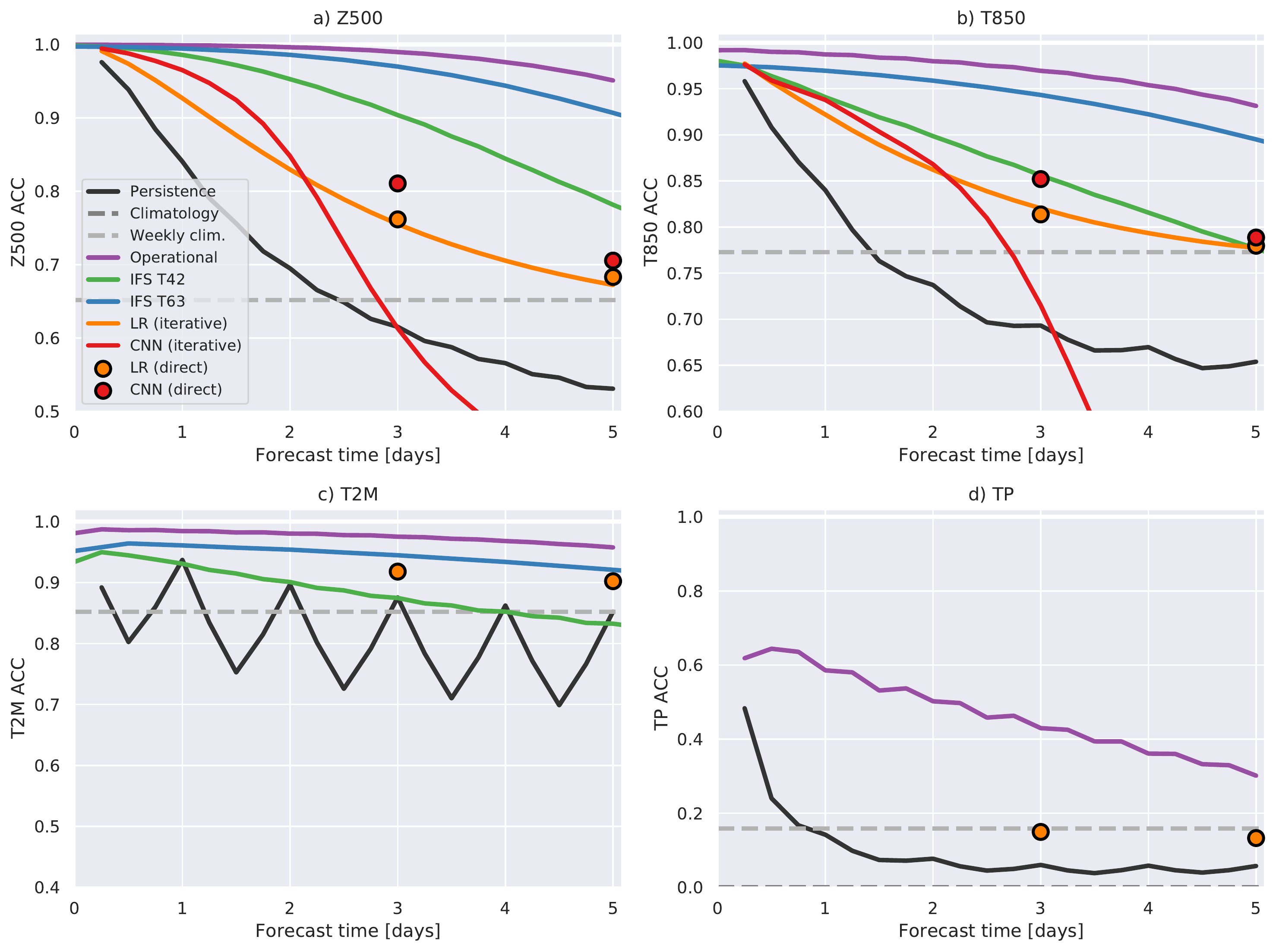}
\caption{ACC of a) 500\,hPa geopotential, b) 850\,hPa temperature, c) 2-meter temperature and d) 6-hourly accumulated precipitation for different baselines at 5.625\textdegree \ resolution.}
\label{fig:acc}
\end{figure*}
\clearpage
\begin{figure*}[th!]
 \includegraphics[width=\linewidth]{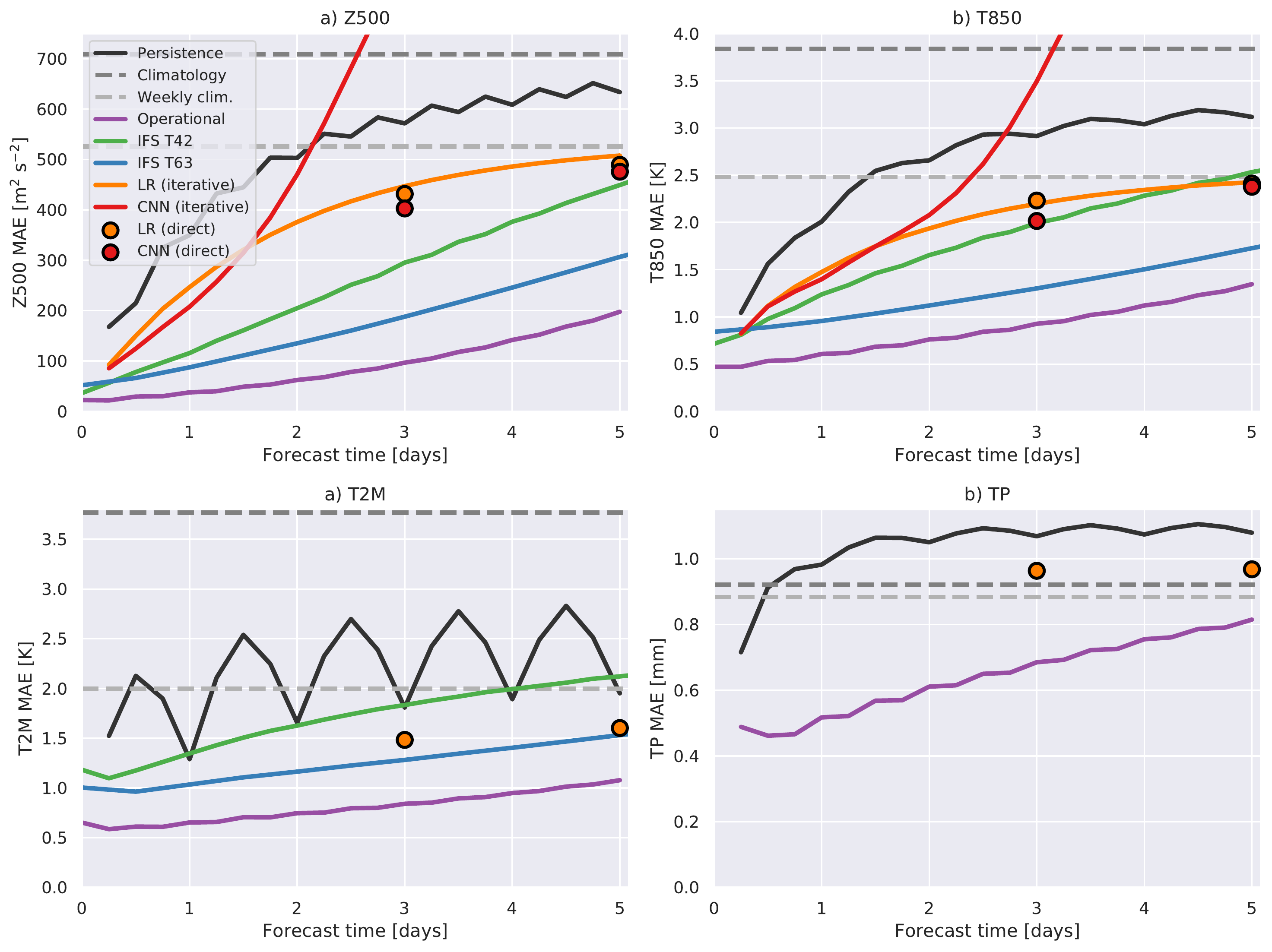}
\caption{MAE of a) 500\,hPa geopotential, b) 850\,hPa temperature, c) 2-meter temperature and d) 6-hourly accumulated precipitation for different baselines at 5.625\textdegree \ resolution.}
\label{fig:rmse}
\end{figure*}
\clearpage
\appendixtables
\begin{table}[th!]
\caption{Baseline ACC for 3 and 5 days forecast time at 5.625\textdegree \ resolution. Best machine learning baseline and physical model are highlighted. TP is 6 hourly accumulated precipitation.}
\begin{tabular}{l||l|l|l|l}
 \multicolumn{4}{c}{\hspace{180pt} ACC (3 days / 5 days)} \\
\textbf{Baseline} & \textbf{Z500} & \textbf{T850} & \textbf{T2M} & \textbf{TP}\\ \hline \hline
Persistence & 0.62 / 0.53 & 0.69 / 0.65 & 0.88 / 0.85 & 0.06/0.06 \\ \hline
Climatology & 0 & 0  & 0 & 0\\ \hline
Weekly climatology & 0.65 & 0.77  & 0.85 & 0.16 \\ \hline
Linear regression (direct) & 0.76 / 0.68 & 0.81 / 0.78  & 0.92 / 0.90 & 0.15 / 0.13\\ \hline
Linear regression (iterative) & 0.76 / 0.67 & 0.82 / 0.78  & & \\ \hline
CNN (direct) & 0.81 / 0.71 & 0.85 / 0.79  & & \\ \hline
CNN (iterative) & 0.61 / 0.41 & 0.72 / 0.31  & & \\ \hline
IFS T42 & 0.90 / 0.78 & 0.86 / 0.78  & 0.87 / 0.83 & \\ \hline
IFS T63 & 0.97 / 0.91 & 0.94 / 0.90  & 0.94 / 0.92 & \\ \hline
Operational IFS & 0.99 / 0.95 & 0.97 / 0.93 & 0.98 / 0.96 & 0.43 / 0.30 
\end{tabular}
\label{tab:scores_acc}
\end{table}

\begin{table}[th!]
\caption{Baseline MAE for 3 and 5 days forecast time at 5.625\textdegree \ resolution. Best machine learning baseline and physical model are highlighted. TP is 6 hourly accumulated precipitation.}
\begin{tabular}{l||l|l|l|l}
 \multicolumn{4}{c}{\hspace{180pt} MAE (3 days / 5 days)} \\
\textbf{Baseline} & \textbf{Z500 [m$^2$ s$^{-2}$]} & \textbf{T850 [K]} & \textbf{T2M [K]} & \textbf{TP [mm]}\\ \hline \hline
Persistence & 572 / 634 & 2.91 / 3.12 & 1.81 / 1.95 & 1.07 / 1.08\\ \hline
Climatology & 708 & 3.84 & 3.77 & 0.92\\ \hline
Weekly climatology & 525 & 2.48 & 2.00 & 0.88\\ \hline
Linear regression (direct) & 431 / 489 & 2.23 / 2.41 & 1.48 / 1.60 & 0.96 / 0.97\\ \hline
Linear regression (iterative) & 447 / 508 & 2.20 / 2.42 & & \\ \hline
CNN (direct) & 403 / 476 & 2.02 / 2.38 & & \\ \hline
CNN (iterative) & 892 / 1263 & 3.49 / 7.49 & & \\ \hline
IFS T42 & 295 / 449 & 1.99 / 2.53 & 1.83 / 2.12 & \\ \hline
IFS T63 & 188 / 307 & 1.30 / 1.73 & 1.28 / 1.53 & \\ \hline
Operational IFS & 97 / 198 & 0.93 / 1.35 & 0.84 / 1.08 & 0.69 / 0.81\\ \hline
\end{tabular}
\label{tab:scores_mae}
\end{table}









\authorcontribution{SR, PD, SS and JW conceived the idea. SR prepared the data and baselines and led the writing. All authors contributed to the manuscript.} 

\competinginterests{The authors declare no competing interests.} 


\begin{acknowledgements}
Stephan Rasp acknowledges funding from the German Research Foundation (DFG). We thank the Copernicus Climate Change Service (C3S) for allowing us to redistribute the data. Peter D. Dueben gratefully acknowledges funding from the Royal Society for
his University Research Fellowship and the ESIWACE2 project. The ESIWACE2 project have received funding from the European Union’s
Horizon 2020 research and innovation programme under grant agreement No 823988.
\end{acknowledgements}







\bibliographystyle{copernicus}
\bibliography{references.bib}

\end{document}